\documentclass[journal]{IEEEtran}
\usepackage{cite}
\ifCLASSINFOpdf
\else
\fi
\usepackage{caption}
\usepackage{graphicx}
\usepackage{subfigure}
\usepackage{color}
\usepackage{arydshln}
\usepackage[noend]{algpseudocode}
\usepackage{algorithm}
\usepackage{multirow}

\begin{document}
\captionsetup[figure]{labelsep=period, singlelinecheck=off} 

\title{Toward 6G Native-AI Network: Foundation Model based Cloud-Edge-End Collaboration Framework }

\author{
    Xiang~Chen,
    Zhiheng~Guo,
    Xijun~Wang,
    Chenyuan~Feng,
    Howard~H.~Yang,\\
    Shuangfeng Han,
    Xiaoyun Wang,
    and Tony~Q.~S.~Quek,~\IEEEmembership{Fellow,~IEEE} 
\thanks{This work has been submitted to the IEEE for possible publication. Copyright may be transferred without notice, after which this version may no longer be accessible.}
\thanks{This work was supported in part by the National Key Research and Development Program of China under Grant 2022YFB2902004, in part by the Open Research Fund from the Guangdong Provincial Key Laboratory of Big Data Computing, The Chinese University of Hong Kong, Shenzhen, under Grant No. B10120210117-OF09, and in part by the National Natural Science Foundation of China under Grants 62271513 and 62301328. (Corresponding authors: Xijun Wang; Xiaoyun Wang.)}
\thanks{Xiang~Chen is with Sun Yat-sen University, China, and also with the Guangdong Provincial Key Laboratory of Big Data Computing, China; Zhiheng~Guo and Xijun~Wang are with Sun Yat-sen University, China; Chenyuan~Feng is with EURECOM, France; Howard~H.~Yang is with Zhejiang University, China; Shuangfeng Han and Xiaoyun Wang are with China Mobile Research Institute, China; Tony~Q.~S.~Quek is with Singapore University of Technology and Design, Singapore.} \vspace{-1.5em}}

\maketitle

\begin{abstract}
Future wireless communication networks are in a position to move beyond data-centric, device-oriented connectivity and offer intelligent, immersive experiences based on multi-agent collaboration, especially in the context of the thriving development of pre-trained foundation models (PFM) and the evolving vision of 6G native artificial intelligence (AI). 
Therefore, redefining modes of collaboration between devices and agents, and constructing native intelligence libraries become critically important in 6G. In this paper, we analyze the challenges of achieving 6G native AI from the perspectives of data, AI models, and operational paradigm. Then, we propose a 6G native AI framework based on foundation models, provide an integration method for the expert knowledge, present the customization for two kinds of PFM, and outline a novel operational paradigm for the native AI framework.
As a practical use case, we apply this framework for orchestration, achieving the maximum sum rate within a cell-free massive MIMO system, and presenting preliminary evaluation results.
Finally, we outline research directions for achieving native AI in 6G.
\end{abstract}

\section{Introduction}
In recent decades, wireless communication networks have undergone remarkable advancements, driven by numerous algorithms enhancing data rate, connection density, etc. Concurrently, the last decade has been known as the ``magic decade" for artificial intelligence (AI), marked by the consistent success of technologies like deep learning (DL) and reinforcement learning (RL) across various domains, shaping a data-driven paradigm for future wireless networks. 
As we approach the 6G era, the focus is shifting towards ubiquitous intelligence through the integration of communication and AI \cite{yy2024}, with AI seamlessly permeating the entire network ecosystem.
Consequently, native AI is poised to usher in a new era of intelligent connectivity, promising to revolutionize how we interact with and benefit from wireless technologies \cite{Gort2025Forecasting, Liu2020Vision, survey2024}.

Recent efforts have paved the way for native-AI networks, marking significant progress in the field. This advancement has manifested in two key directions. Firstly, communication networks have begun incorporating a substantial number of well-trained AI models to enhance quality of service/experience (QoS/QoE). Simultaneously, there has been a paradigm shift where communication networks are being integrated as an intrinsic part of AI systems, improving overall performance \cite{dj2024}. This transition from AI for Communication Network to Communication Network for AI is illustrated in \cite{Wen2022Nine}. 
Additionally, the advent of Large Language Models (LLMs), exemplified by ChatGPT, has spurred enormous innovations. 
Compared with traditional AI networks, their unique capabilities, (e.g., natural language interaction, multimodal data handling) are crucial for enhancing the performance and interactivity of 6G networks. Notably, several efforts have been made to integrate LLMs into 6G, leading to the development of communication-oriented LLMs \cite{Du2025Mixture, Zirui2023Big} and the design of novel wireless network architectures \cite{Yuxuan2023NetGPT}.
These aforementioned works fall primarily into two classes: The first involves training task-oriented AIs tailored for specific functions, whereas in the second, an LLM is used to address the full spectrum of requirements in 6G. However, both approaches exhibit notable limitations, especially when dealing with limited computing and storage resources. The former approach lacks generalizability and necessitates a substantial retraining overhead for new tasks, while the latter encounters challenges when deployed at the edge due to its resource-intensive nature.

Given these challenges, a fundamental question arises: How can native AI be realized in 6G, particularly with a foundation model that can handle more than just linguistic data types? To answer this question, this paper proposes a native AI framework built upon a foundation model and powered by cloud-edge-end collaboration. Specifically, we begin by analyzing the prerequisites and challenges associated with the integration of communication and AI from three perspectives: data, AI models, and operational paradigms. 
We then discuss how AI is deployed and works in a cloud-edge-end collaboration and emphasize existing limitations. 
In this context, we fully leverage the capabilities of pre-trained foundation models (PFM)-based agents to process multimodal data (e.g., received signals, textual information) and make decisions on request identification, task planning, and tool invocation, etc.
Based on these capabilities, we establish a 6G-native AI framework comprising three key components: a communication-specific PFM, expert knowledge integrated through advanced retrieval-augmented generation (RAG), and hierarchical multi-agent collaboration.
To demonstrate the framework's potential, we present a use case focusing on network orchestration, illustrating its substantial contributions to 6G networks. Finally, we outline essential research directions required to fully harness the potential of native AI in 6G.

\section{Research Challenges in 6G native AI} \label{sec: Challenges}
\subsection{Data Perspective} \label{subsec: Data Chanllenges}
Leveraging vast datasets for training is crucial to ensuring optimal QoS/QoE in AI-driven communication networks.
Despite the existence of AI data pipelines for collection, processing, storage, and retrieval, 6G networks face significant challenges. 
Hereafter, we scrutinize three unique data categories in 6G to enhance enhancing data volume and efficiency:
\subsubsection{Real-world Data} 
Real-world data acquired from the air interface in 6G networks constitutes a fundamental resource for training AI models and is indispensable for advancing emerging applications, particularly in integrated sensing and communication scenarios.
However, the data collection in 6G networks presents significant challenges, including ensuring data quality, managing labeling costs, and navigating restrictions on privacy-sensitive information. Furthermore, compared with computer vision and natural language processing, data collected from air interface may not be human-interpretable, posing significant challenges for labeling.
To overcome these hurdles, it is crucial to foster a coordinated approach to data collection, sharing, and utilization while respecting privacy concerns and regulatory requirements.

\subsubsection{Synthetic Data}
Characterized by its well-defined structure and ease of acquisition and labeling, synthetic data offers an effective solution to the challenges of real-world data collection. 
In 6G networks, combining generative AI (GenAI) with communication networks enables the generation of large volumes of data that follow the same distribution as real-world data, while maintaining low operational costs. In this sense, it is a promising technique in 6G for enhancing the diversity and richness of synthetic data. Additionally, to address data labeling challenges, the introduction of digital twins technology in 6G has emerged as an effective solution, establishing itself as one of the key sources of synthetic data.
However, the key challenge still lies in creating virtual simulated environments that closely resemble real-world conditions.

\subsubsection{Standards and Technical Reports}
Communication networks are substantially constrained by data processing requirements, such as 3GPP-defined message structures. These standards serve as crucial domain knowledge, enhancing the inference capabilities of communication-specific PFMs. Additionally, technical reports and white papers offer detailed technical information, particularly valuable when implementation guidelines are lacking in the standards. While these resources are usually publicly available and easily accessible, two main challenges remain: constructing effective representations of the complex information contained in these documents, and developing efficient methods for searching and utilizing relevant information across standards and technical reports.

\vspace{-1.2em}
\subsection{AI Model Perspective} \label{subsec: AI Chanllenges}
PFMs, such as BERT and GPT, have amassed extensive knowledge across various domains and achieved remarkable success in Computer Vision (CV) and Natural Language Processing (NLP). This success has led to the development of numerous significant applications, demonstrating their efficacy in addressing human needs. 
However, applying PFMs to communication networks poses challenges. These models need require meticulous customization to incorporate domain-specific knowledge. Moreover, deploying an entire PFM on resource-constrained devices is often impractical, requiring lightweight, task-oriented AI models. Despite enormous strides in enhancing AI for 6G networks, the industry lacks standardized task-oriented models to address practical challenges. To overcome this, native AI should integrate two types of intelligence:

\subsubsection{Communication-specific PFM} 
The exceptional semantic recognition, learning capabilities, and inference prowess of PFMs make them a viable option for efficient intent recognition, task partitioning, and fine-grained resource allocation. These attributes enable a novel orchestration mechanism, crucial for managing the diverse and heterogeneous task requests expected in 6G. However, leveraging the potential of PFMs and developing a communication-specific PFM that can effectively handle various tasks remains a significant challenge. 

\subsubsection{Task-oriented AI Toolkit} 
Deploying PFMs at the network edge or on end-user devices is challenging due to their high resource demands. Lightweight AI models offer a promising alternative, aligning with 6G's vision of pervasive intelligence. As a result, task-oriented AI models are set to become key tools in 6G network development. To maximize their potential, a standardized task-oriented AI toolkit is essential to enhance network efficiency and intelligence.
However, to ensure the models remain effective and relevant {{as network conditions and user demands change}}, ongoing maintenance and updates of this toolkit are crucial for long-term deployment of communication networks.

\begin{figure*}[t]
\centerline{\includegraphics[width=0.95\textwidth]{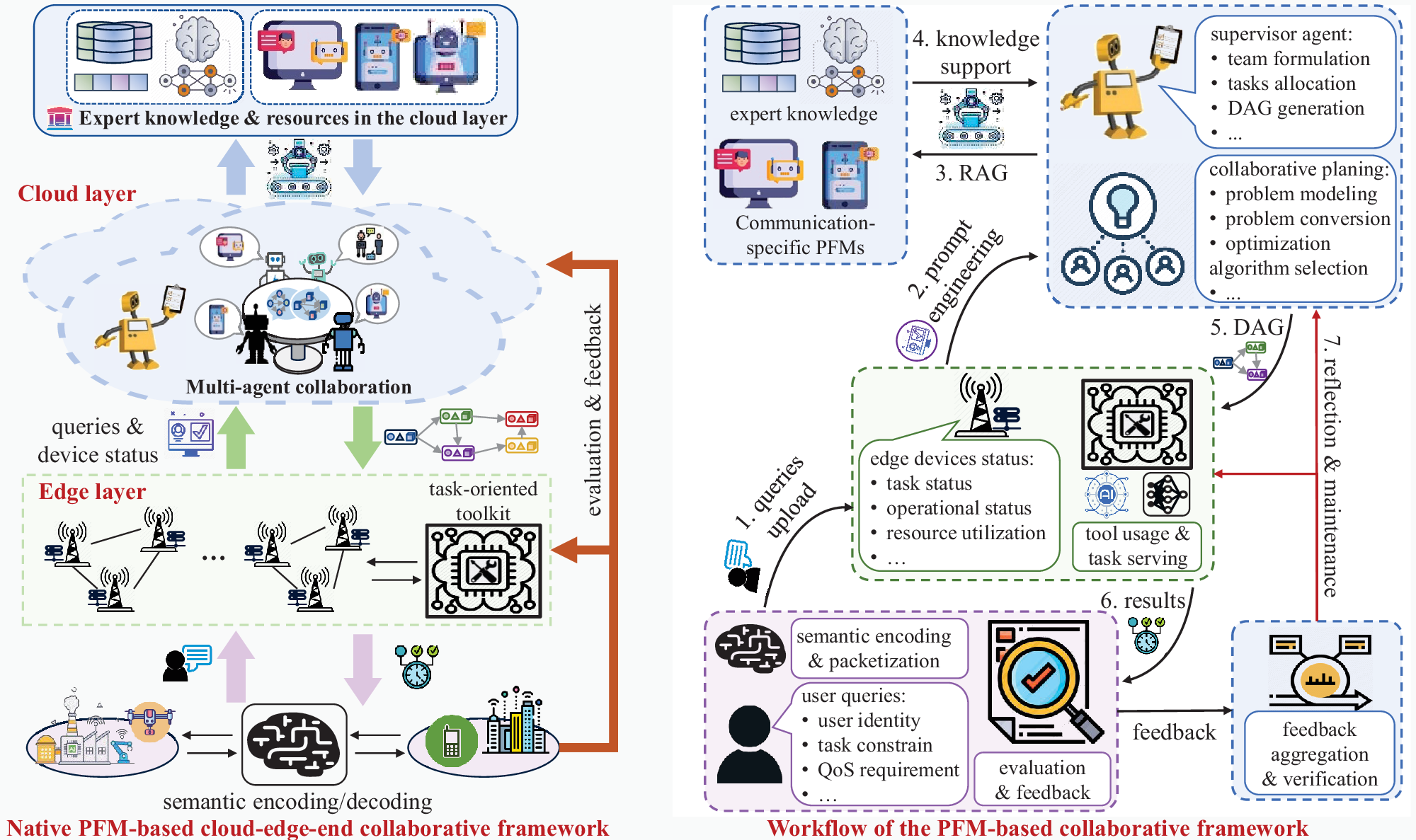}}
\caption{The proposed native AI framework with cloud-edge-end collaboration.}
	\label{fig: Framework}
    \vspace{-1em}
\end{figure*}
\subsection{Operational Perspective} \label{subsec: Work Paradigm Chanllenges}
LLMs, supported by extensive knowledge, have demonstrated impressive capabilities in addressing real-world problems. However, when relevant knowledge is missing or unreliable, LLMs often provide overconfident responses, leading to hallucinations and biases.
Addressing communication tasks typically involves multi-step processes and requires a wide range of continuously evolving interdisciplinary knowledge, including information theory, optimization, etc.
While current LLMs, operating on the one-step input-output execution mode, struggle to synthesize information from various sources and formats.
This results in persistent knowledge gaps for LLMs applied to communication systems, and exacerbates the issue of ``context dilution", where critical data is obscured by information overload.
Therefore, communication-specific PFMs should incorporate two innovative operational paradigms:

\subsubsection{Task Planning Mode}
Task planning mode in communication-specific PFMs addresses the strict procedural standards and specific constraints inherent in communication tasks. By leveraging domain knowledge, PFMs can strategically plan task procedures, including user intent recognition, task segmentation, and resource allocation. This approach enables PFMs to utilize tools and serve each subtask more efficiently. However, the primary challenge lies in designing a workflow tailored to the communication field that can effectively implement this task planning mode, ensuring optimal performance within the unique constraints of communication systems.

\subsubsection{Multi-agent Collaboration}
Multi-agent collaboration presents a paradigm shift from single AI agent to an interconnected framework, offering an intuitive and efficient way to achieve complementary knowledge in communication tasks. By enabling multiple LLMs with domain-specific knowledge to collaborate, this approach facilitates knowledge sharing, alleviates the issue of ``context dilution" and provides more systematic solutions. However, how to develop a method for dynamically assembling multi-LLM collaboration teams, along with clearly define the specific responsibilities of each LLM are the pressing issues to enable the collaboration.

\vspace{-0.2em}
\section{Native AI Framework for 6G Networks}  \label{sec: Framework}
In this section, we propose a native AI framework for 6G networks, and delve into three key technologies that underpin and enable the framework's functionality.

\vspace{-1em}
\subsection{Architecture of the Native AI Framework}  \label{subsec: Architecture and Workflow of the Native AI Framework}
{{
In this paper, 6G-native AI integrates PFM-based agents and task-oriented AIs. PFM-based agents demand substantial computational and memory resources to deliver advanced global reasoning capabilities, making them ideal for deployment on cloud nodes equipped with ample resources. Conversely, task-oriented AIs are optimized for efficient operation under resource constraints, enabling specialized service processing, and better suited for deployment on edge nodes to independently handle latency-sensitive tasks locally.
}}
As shown in Fig.~\ref{fig: Framework}, we build a native AI framework that leverages cloud-edge-end collaboration. In the cloud, we integrate the expert knowledge using a vector database (VB) and knowledge graphs constructed by advanced RAG. We also build an AI resource pool comprising a set of communication-specific PFMs.
These elements, along with traditional communication resources, serve as key components of the available system resources.
The edge is equipped with a task-oriented toolkit that includes task-oriented small AI models and validated, mature algorithms for wireless communication \cite{Kartsakli2023An}. 
End-user devices are fitted with lightweight AI model, enabling them to encode user intent and feedback into semantic information, providing key information for task generation.

The workflow of our framework begins with task initiation. End-user devices initiate tasks by semantically encoding their queries. These queries, along with the edge's status information, are processed through prompt engineering. The PFM then generates multiple relevant queries that align with the user's original intent. These generated queries, together with the original query, are transmitted to the agent as context for knowledge retrieval. This approach helps eliminate the loss of global retrieval information caused by incomplete or ambiguous user queries. Subsequently, expert knowledge is filtered, retrieved from the database, and forwarded to the agents for further processing.

To generate responses, our framework implements a hierarchical multi-agent collaboration scheme, comprising a supervisor agent and specialist agents. The supervisor agent identifies the user's intent, breaks the task into subtasks, and allocates them to specialist agents based on their specific functionalities. Each specialist agent orchestrates its assigned subtask, determining required resources and feasible algorithms. The supervisor agent then reviews these results, deciding whether to accept or request revisions, and ultimately generates the task Directed Acyclic Graph (DAG) for execution at the edge.

Upon receiving the task DAG, the edge selects the appropriate models from the toolkit based on the DAG's instructions. It then either executes the task with the end-user device's data or deploys the models to the device for local execution.
After task completion, user-end devices provide feedback to the cloud. Given the subjective nature of this feedback and potential outliers, the supervisor agent aggregates and filters it before activating the reflection mechanism. 

\begin{figure}[t]
\centerline{\includegraphics[width=0.4959\textwidth]{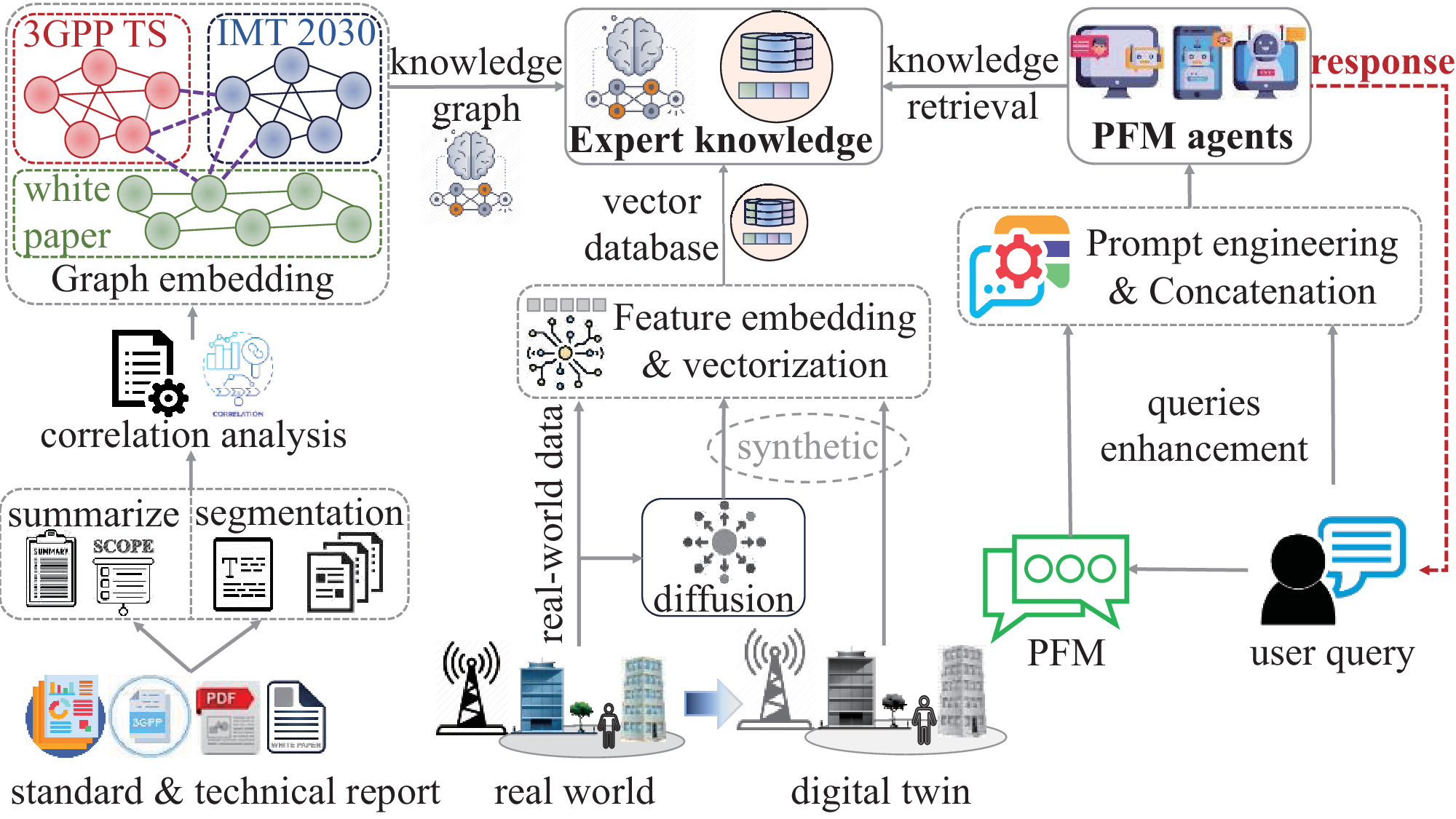}}
	\caption{An illustration of the advanced RAG.}
	\label{fig: Enhanced RAG}
    \vspace{-1em}
\end{figure}
\vspace{-1em}
\subsection{Key Technologies}  \label{subsec:Key Technologies}
\subsubsection{Advanced RAG} \label{subsubbsec: Advanced RAG}

The advent of RAG provides a solution for integrating expert knowledge into PFMs shown in Fig.~\ref{fig: Enhanced RAG}. We employ a vanilla RAG \cite{Gao2024Retrieval} to index real-world and synthetic data, with the latter generated via digital twins and GenAI (e.g., diffusion models). The process begins with data preparation, which involves cleaning and extracting raw data from various formats such as MAT and JSON, and then converting it into a unified format. 
These data are then enriched with metadata, including communication environments, task descriptions, and other relevant contexts, to enhance their utility. Once processed, the data are encoded into embedding vectors and stored in a VB for efficient retrieval.

For standards and technical reports, we introduce a graph-based RAG to capture the relationships among various data, enhancing the PFM’s ability to address comprehensive queries.
Unlike the aforementioned VB, the graph-based RAG constructs a graph database whose entities are knowledge graphs derived from standards and reports. This representation consists of a triplet: graph weights, edge weights, and node weights.
Specifically, the graph weight represents the summary descriptions of standards and reports (e.g., Scope/Introduction in 3GPP standards), including their target scenarios, tasks type, and functionalities. Standards and reports with identical or highly similar summary descriptions are grouped into the same knowledge graph. 
To further obtain the edge and node weights, we first convert the multimodal data into text by tokenization and divide them into multiple chunks. The edge weights represent the relationships between these chunks, determined by contextual connections, mutual citation, and the frequency of cross-references between the corresponding chunks. Node weights are represented by the semantic information of each chunk.
Based on the definition of the triplet, we can generate the corresponding embedding graph using off-the-shelf graph database management systems like modular GraphRAG system \cite{Edge2024From}, or custom-built graph neural networks.

Expert knowledge retrieval is achieved through a combination of semantic similarity-based retrieval and subgraph matching.
We begin by generating several related queries based on the original user query, effectively transforming a single query task into multiple related query tasks. 
These generated queries undergo semantic encoding, similar to the data stored in the VB, enabling relevant knowledge to be retrieved through semantic similarity matching from the database. The retrieved knowledge is then fetched, deduplicated, and re-ranked, with the top-$K$ results forwarded to the agent.
Additionally, we generate a graph query based on these queries. In this graph query, the graph weight represents the summary description of the original user query, the edge weights reflect the relationships between the queries, and the node weights represent the semantic information. Subsequently, we identify subgraphs in the graph database that match this particular pattern and retrieve nodes and relationships along specific paths connecting these subgraphs. With these retrieved expert knowledge, we can enhance the PFM's responses and alleviate hallucination by grounding its outputs in verified information.

\subsubsection{Agentic Workflow} \label{subsubsec: Autonomous Working Mode}
An agentic workflow is a self-directed, proactive approach to task management that emphasizes autonomy, initiative, and strategic thinking. 
This methodology enables native AI to manage work processes and outcomes in 6G networks, focusing on three key aspects:

\textit{Planning:} 
In our proposed framework, PFMs transcend their traditional operational paradigm of simple query-response generation. By incorporating prompt-based planning capabilities, these PFMs are enhanced to function as autonomous agents, capable of decomposing complex communication tasks into hierarchical, manageable subtasks and generating structured action sequences for their resolution. This planning mechanism results in the formation of a DAG of tasks. 
Each subtask is comprehensively defined through textual descriptions or corresponding prompts, encompassing crucial orchestration parameters. These parameters include subtask identification and description, resource constraints, predefined evaluation criteria, tool selection, and execution node specifications.

\textit{Tool usage:} 
The framework enables agents to leverage a diverse toolkit for task execution. Upon receiving orchestrated subtask specifications, agents dynamically select appropriate algorithmic tools and determine optimal execution nodes (e.g., specific edge nodes) based on the given parameters. Notably, when the algorithms are not available in the existing toolkit, agents can utilize code generation capabilities to programmatically create new algorithms, drawing upon retrieved expertise.

\begin{figure*}[t]
\centerline{\includegraphics[width=0.9\textwidth]{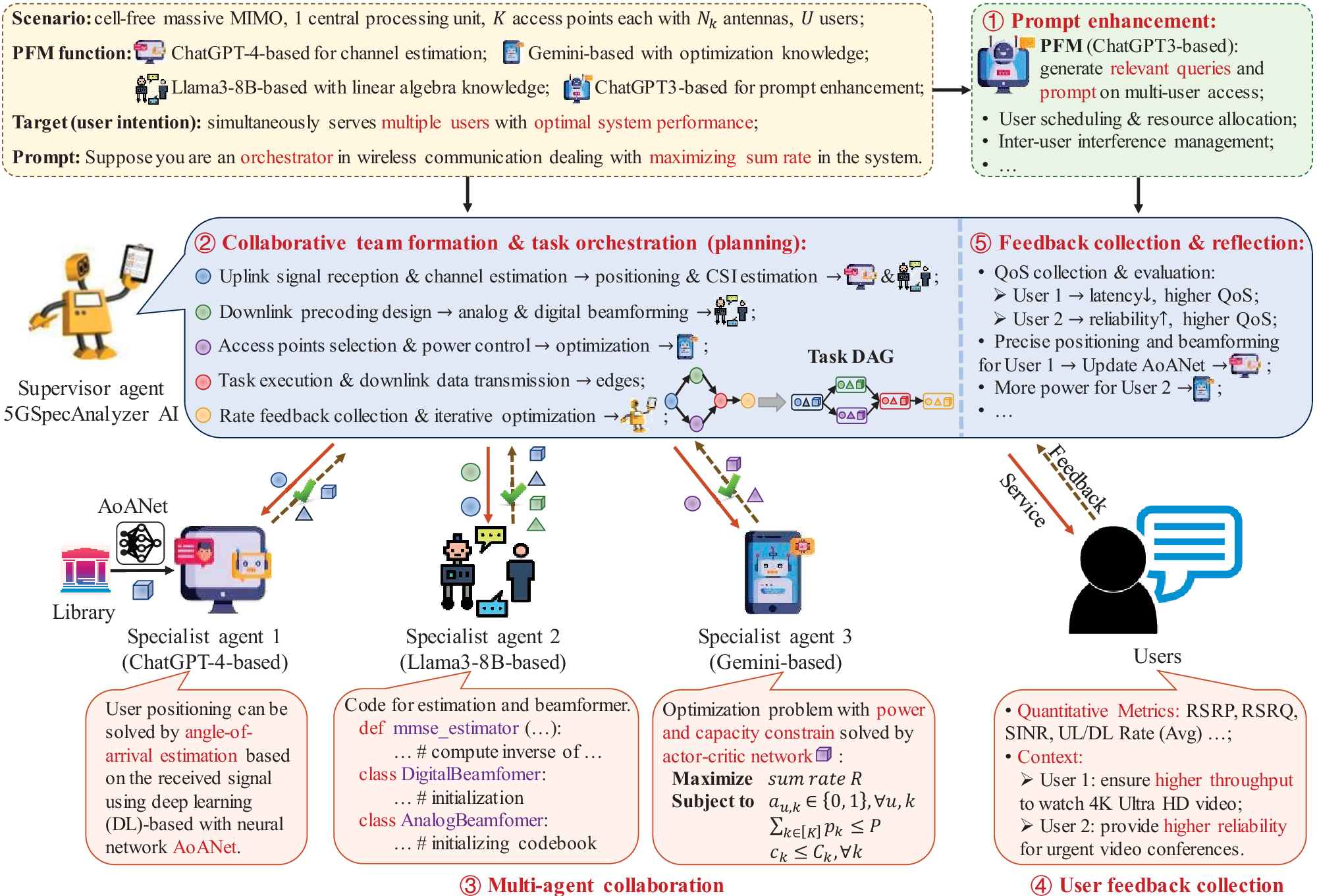}}
\caption{A case study of task orchestration by multi-agent collaboration.}
\label{fig: Simulation}
\vspace{-1em}
\end{figure*}

\textit{Reflection:}
Reflection is a key technology that enables PFMs to adapt to diverse tasks and achieve continuous self-improvement. Our framework implements two distinct types of reflection mechanisms.
The first is self-reflection, where PFMs conduct comprehensive self-assessment of their planning steps and proposed solutions before finalizing the task DAG. 
This proactive approach involves predicting potential execution outcomes, allowing for optimization of the task orchestration before actual deployment. 
The second is feedback-based reflection, where PFMs systematically analyze real-world execution results. This retrospective analysis allows PFMs to assess areas for improvement in their orchestration strategies, learning from both successes and failures. The insights gained from this process inform future planning decisions, creating a continuous improvement loop. 
{{
This kind of feedback primarily comprises quantitative metrics of communication system (e.g., SINR, RSRP) and users' natural language feedback (see Fig.~\ref{fig: Simulation}).
The former, widely used in wireless communication systems, provides real-time performance evaluations. While the latter captures users' subjective perceptions by directly measuring their QoE. With this feedback, the PFMs can more intuitively comprehend user requirements.
}}
These PFMs can be updated by parameter-efficient fine-tuning methods, e.g., low-rank adaptation \cite{Ning2023Parameter} and zeroth-order gradient update \cite{Tang2024Zeroth}.

\begin{figure}[t]
\centerline{\includegraphics[width=0.4959\textwidth]{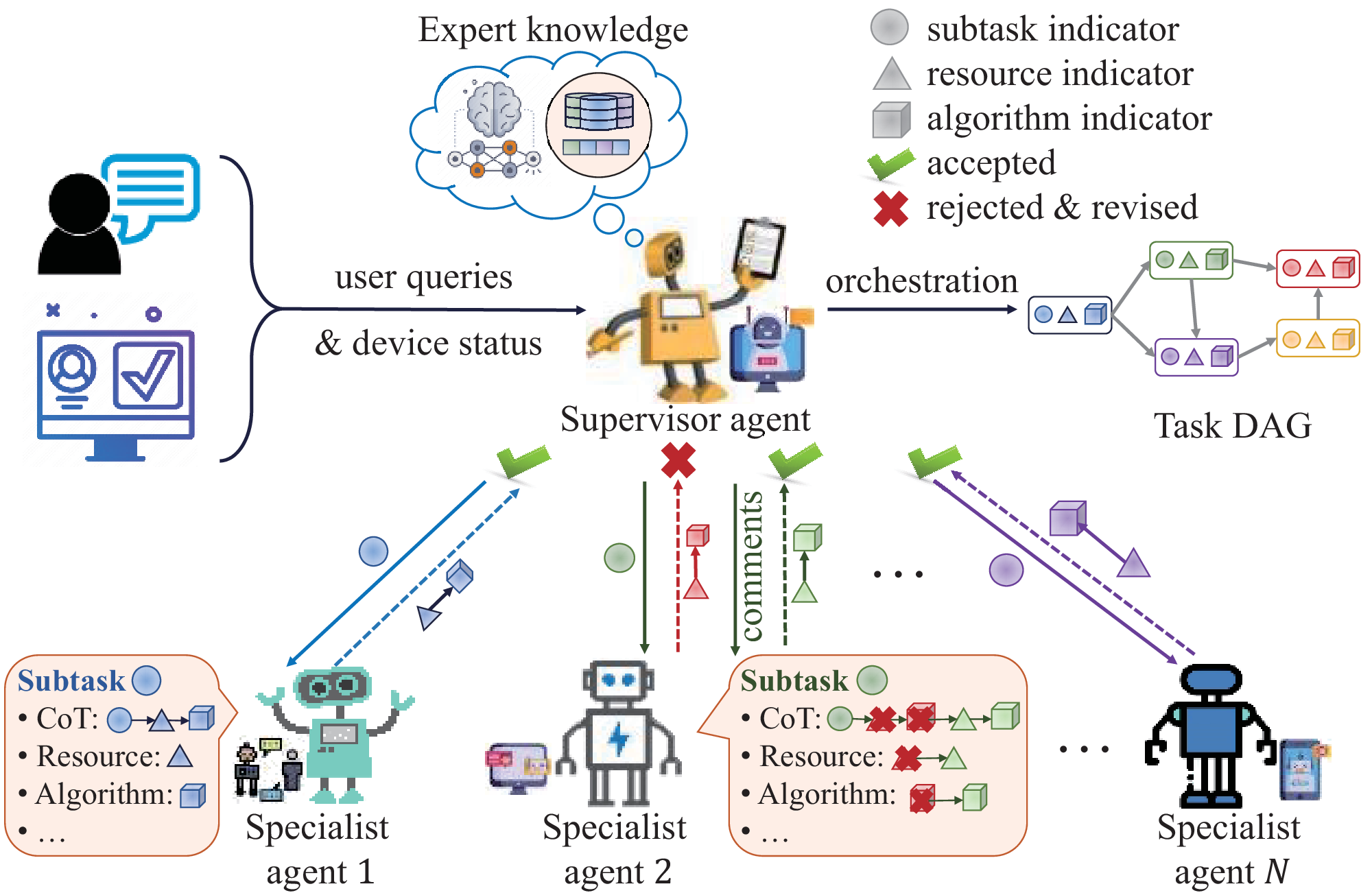}}
	\caption{An illustration of hierarchical multi-agent collaboration.}
	\label{fig: Hierarchical Multi-agent Collaboration}
    \vspace{-1em}
\end{figure}

\subsubsection{Hierarchical Multi-agent Collaboration} \label{subsubsec: Hierarchical Multi-agent Collaboration}
To address the diversity of user queries and accommodate time-varying environment, our framework implements a hierarchical multi-agent collaboration mechanism, which overcomes the limitations of single-agent architectures by integrating supervisor and specialist agents. The supervisor agent, serving in a supervisory capacity, identifies user intent, decomposes complex tasks, and reviews specialist agents' plans. Conversely, specialist agents are optimized for specific task domains and orchestrate their assigned subtasks with domain-specific expertise.

Our proposed hierarchical multi-agent collaboration mechanism is illustrated in Fig.~\ref{fig: Hierarchical Multi-agent Collaboration}. The supervisor agent employs contextual information and retrieved knowledge to identify user intentions and decompose the task into subtasks, each defined by specific parameters including description, resource constraints, and predefined criteria. These subtasks are then mapped into a DAG to represent their interdependencies. The supervisor agent proceeds to dynamically assemble an appropriate team of specialist agents, with selection based on their functional descriptions and alignment with task requirements.

{{
In the execution phase, 
the supervisor agent decomposes the communication task, represented by user request, into specific communication subtasks and delegates subtasks to corresponding specialist agents.
The specialist agents then perform task planning by applying domain knowledge. Note that if multiple specialist agents are assigned to collaboratively complete the same subtask, they will exchange information to propose a joint solution. Otherwise, if each specialist agent is allocated an independent subtask, it is solely responsible for executing its assigned subtask without exchanging information with other specialist agents.
The specialist agents will finally selects appropriate algorithms from the toolkit and report the corresponding solution to the supervisor agent for evaluation.
The supervisor agent assesses these plans against predefined criteria. Plans that fail to meet requirements are returned with highlighted deficiencies for revision and resubmission.
Upon successful acceptance of all plans, the supervisor agent generates a comprehensive task DAG, which encompasses detailed subtask instructions, resource allocations, and algorithm selection, ultimately transmitting this orchestration plan to the designated execution node.
Finally, once the framework receive the feedback from users, the supervisor agent determine the necessary update for specialist agents.
}}

\begin{figure}[h]
	\centering{ 
	\vspace{-0.35cm} 
	\subfigtopskip=1pt 
	\subfigbottomskip=1pt 
	\subfigcapskip=-5pt 
    \subfigure[Average of the sum rate]{
	\label{subfig: AveSumRate}
	\includegraphics[width=0.4\textwidth]
    {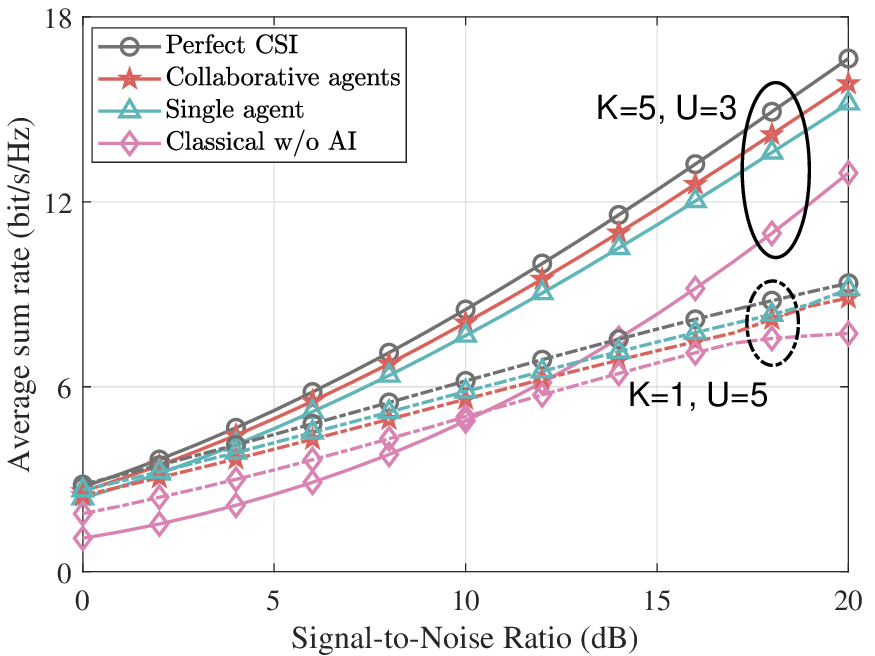}} \\
    \subfigure[CDF of the sum rate]{
	\label{subfig: SumRateCDF}
	\includegraphics[width=0.4\textwidth]
    {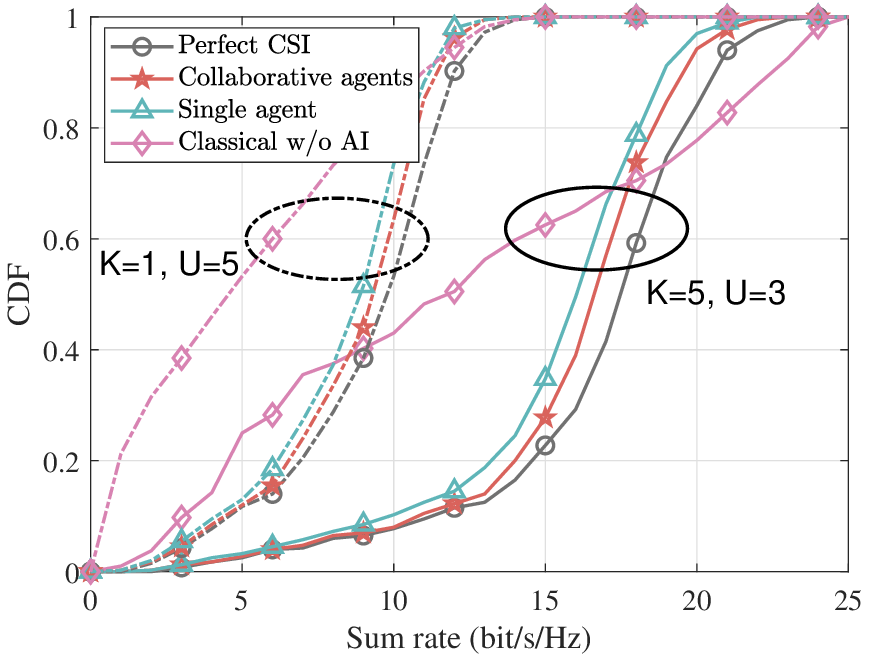}}
    }
	\caption{Performance comparison between our proposed framework with collaborative agents and three baseline schemes.} 
	\label{fig: sum rate evaluation}
    \vspace{-1em}
\end{figure}

\vspace{-0.8em}
\section{Case Study: Task Management and Orchestration} \label{sec: Case Study}
{{
In this section, we address a multi-user access in a cell-free massive MIMO system with a single central processing unit (CPU).
Two scenarios are considered: one with $K=5$ access points (AP) and $U=3$ end-user devices, and another with $K=1$ and $U=5$.
}} Each AP has a uniform linear array of $N_t=64$ antennas, while each end-user device has a single antenna. The channel between each AP and the end-user device is modeled as a block fading channel, mainly determined by the line-of-sight propagation path. The downlink sum rate serves as a system performance metric.  

The supervisor agent is implemented by ``5GSpecAnalyzer AI" \cite{Archana2024Spec}, providing expert analysis based on 3GPP standards. Specialist agents are customized from existing LLMs through prompt fine-tuning, with prompts generated by the supervisor agent during task orchestration. The specialist agent configuration includes: a \mbox{GPT-4}-based PFM for angle-of-arrival (AoA) estimation tasks, a Gemini-based PFM with optimization theory expertise, and a Llama3-8B-based PFM specialized in linear algebra and precoding implementations. To improve RAG quality, the system incorporates GPT-3 for query enhancement. The ``5GSpecAnalyzer AI" additionally functions as a critic PFM facilitating the reflection mechanism.
 
The task orchestration process, shown in Fig.~\ref{fig: Simulation}, begins with providing system scenario specifications and PFM functional descriptions as foundational knowledge to the supervisor agent. The primary task objective of optimizing system performance for multiple users undergoes query enhancement through GPT-3, generating multiple related prompts which, in conjunction with the original query, are transmitted to the supervisor agent. Utilizing these enhanced prompts, the supervisor agent decomposes the overall task into ordered subtasks, allocating each to appropriate specialist PFMs. Following task allocation, each specialist PFM executes its specialized function to contribute to the overall solution. The GPT-4-based PFM proposes an advanced DL method for user positioning by AoA estimation. Concurrently, the Llama3-8B-based PFM develops and generates optimized implementation code for channel estimation and precoding algorithms. The Gemini-based PFM, leveraging its expertise in optimization theory, formulates an optimization problem addressing AP selection and downlink transmission strategies.
{{
Upon receiving user feedback, the framework triggers a reflection mechanism, where the supervisor agent determines the necessary updates for the corresponding specialist agents, such as one/few-shot fine-tuning or refining tool invocation. Subsequently, the framework reprocesses the user request after the reflection and delivers the refined result to the user.
}}

In Fig.~\ref{fig: sum rate evaluation}, we present a performance comparison of the proposed framework 
with three baseline schemes. Specifically, our proposed framework leverages multi-agent collaboration to sequentially address DL-based AoA estimation, channel estimation, beam selection, and digital beamforming tasks. 
{{
In contrast, the single agent approach, implemented by ChatGPT-4o for its comprehensive capabilities (e.g., reasoning, mathematical expertise, code generation), autonomously identifies user requests, orchestrates tasks, among others. However, it struggles with task decomposition and often relies on a single module to perform multiple functions, such as adopting an end-to-end hybrid analog-digital beamforming scheme.
}}
The other two baseline schemes, a perfect CSI scenario and a classical approach without AI, use the same beamforming strategy as our framework but rely on perfect CSI and traditional channel estimation techniques, respectively.
{{ The results illustrated in Figs.~\ref{subfig: AveSumRate} and \ref{subfig: SumRateCDF} demonstrate that, in both scenarios, PFM-based schemes achieve superior performance and enhanced stability compared to traditional approaches. Notably, our multi-agent collaborative framework performs more closely to the perfect CSI scenario than the single agent method, substantiating the effectiveness of collaborative agent-based orchestration for optimizing communication systems.
}}

\vspace{-0.5em}
\section{Future Research Directions} \label{sec: Open Issues}
This section outlines key open issues and potential research directions for intelligent communication networks.

\vspace{-1.2em}
\subsection{Evaluation for the Communication-specific PFM}
Evaluating communication-specific PFMs in 6G is challenging due to the diverse tasks they coordinate, requiring comprehensive performance metrics in multiple dimensions. These metrics must address robustness of decision-making, adaptability to dynamic conditions, and efficiency of AI-enhanced functions. Although technical metrics are crucial, achieving high QoE also requires effective human-in-the-loop evaluation, demanding standardized and quantitative methodologies for user feedback. The complexity of linking task performance to overall effectiveness highlights the need for unified evaluation frameworks that assesses PFMs within 6G.

\vspace{-1.2em}
\subsection{Tradeoff between Cost and Performance}
Integrating communication-specific PFMs into communication systems presents a fundamental paradox: the conflict between the resource limitations of the communication infrastructure and the high performance demands of AI models. 
This tension is particularly evident in multi-agent collaboration, where distributed agents for improved performance creates significant communication overhead. This complex interplay of factors leads to ``Impossible Trinity" in AI-communication integration: the seemingly unattainable goal of simultaneously optimizing costs, maintaining operational efficiency, and maximizing performance. As communication systems evolve, addressing this trilemma becomes crucial for native-AI networks deployment that can balance these competing objectives.

\vspace{-1em}
\subsection{Intelligent Communication Standardization}
In wireless communication systems, adherence to standardized operational procedures is crucial, requiring all devices and servers to process signals and deliver services according to established protocols. In contrast, AI technologies are often developed as open-source, establishing de facto standards. 
As a result, ensuring the effective integration and interoperability of AI systems, born within open-source ecosystems, with 6G networks following traditional communication standards is a key and inherently complex challenge.
Although organizations like 3GPP and O-RAN Alliance are beginning to embrace AI, achieving this integration requires close collaboration among standardization bodies, open-source communities, and industry stakeholders to create a unified framework.

\vspace{-0.7em}
\section{Conclusion}
This article presents a transformative paradigm for 6G networks via PFM-based native AI integration. We highlight key challenges in realizing this vision, such as seamless expert knowledge integration, diversity PFM customization, and the comprehensive PFM-based 6G frameworks development.
To address these challenges, we have leveraged cutting-edge technologies, such as advanced RAG, multi-agent collaboration, and PFM reflection mechanisms. A case study was proposed to demonstrate the efficacy of our approach in maximizing sum rates in a multi-access scenario. Finally, we discuss implementation barriers and suggest promising future research directions for realizing native AI in 6G networks.

\vspace{-0.7em}
\bibliographystyle{Reference/IEEEtran}
\bibliography{Reference/Reference}


\vspace{-13mm}

\begin{IEEEbiographynophoto}
{Xiang Chen} (chenxiang@mail.sysu.edu.cn) received the B.E. and Ph.D. degrees from Tsinghua University in 2002 and 2008, respectively. Currently, he is a Full Professor at Sun Yat-sen University, China. 
\end{IEEEbiographynophoto}

\vspace{-13mm}

\begin{IEEEbiographynophoto} 
{Zhiheng Guo} (guozhh7@mail2.sysu.edu.cn) received the B.E. degree from Sun Yat-sen University, China, in 2019, where he is currently pursuing the Ph.D. degree.
\end{IEEEbiographynophoto}

\vspace{-13mm}

\begin{IEEEbiographynophoto}
{Xijun Wang} (wangxijun@mail.sysu.edu.cn) received the B.S. degree from Xidian University, China, in 2005, and the Ph.D. degree from Tsinghua University, China, in 2012. Currently, he is an Associate Professor at Sun Yat-sen University, China. 
\end{IEEEbiographynophoto}

\vspace{-13mm}

\begin{IEEEbiographynophoto}
{Chenyuan Feng} (Chenyuan.Feng@eurecom.fr) received the Ph.D.\ degree from Singapore University of Technology and Design in 2021. Currently, she is a researcher at Eurecom, France and a Marie Skłodowska-Curie Scholar.
\end{IEEEbiographynophoto}

\vspace{-13mm}

\begin{IEEEbiographynophoto}
{Howard H. Yang} (haoyang@intl.zju.edu.cn) received the Ph.D. degree from Singapore University of Technology and Design in 2017. Currently, he is an assistant professor at Zhejiang University, China.
\end{IEEEbiographynophoto}

\vspace{-13mm}

\begin{IEEEbiographynophoto}
{Shuangfeng Han} (hanshuangfeng@chinamobile.com) received the M.S. and Ph.D. degrees from Tsinghua University in 2002 and 2006, respectively. Currently, he is a Principal Researcher of China Mobile Research Institute. 
\end{IEEEbiographynophoto}

\vspace{-13mm}

\begin{IEEEbiographynophoto}
{Xiaoyun Wang} (wangxiaoyun@chinamobile.com) is the Chief Scientist and vice CTO of China Mobile. She is the recipient of multiple National Science and Technology Progress Awards.
\end{IEEEbiographynophoto}

\vspace{-13mm}

\begin{IEEEbiographynophoto}
{Tony Q. S. Quek} (tonyquek@sutd.edu.sg) received the Ph.D. degree from the Massachusetts Institute of Technology in 2008.  Currently, he is the Cheng Tsang Man Chair Professor at Singapore University of Technology and Design, Singapore. He is IEEE Fellow, WWRF Fellow, Fellow of the Academy of Engineering Singapore, and the AI on RAN Working Group Chair in AI-RAN Alliance. 
\end{IEEEbiographynophoto}

\end{document}